\documentclass[fleqn,usenatbib]{mnras}

\usepackage{newtxtext,newtxmath}

\usepackage[T1]{fontenc}

\DeclareRobustCommand{\VAN}[3]{#2}
\let\VANthebibliography\thebibliography
\def\thebibliography{\DeclareRobustCommand{\VAN}[3]{##3}\VANthebibliography}

\usepackage{graphicx}	
\usepackage{amsmath}	

\title[Missing Data Imputation and Photo-$z$ for CSST]{Imputation of Missing Photometric Data and Photometric Redshift Estimation for CSST}

\author[Z. Luo et al.]{
Zhijian Luo$^{1}$\thanks{E-mail: zjluo@shnu.edu.cn},
Zhirui Tang$^{1}$,
Zhu Chen$^{1}$,
Liping Fu$^{1}$,
Wei Du$^{1}$,
Shaohua Zhang$^{1}$,
Yan Gong$^{2,3}$,
\newauthor
Chenggang Shu$^{1}$,
Junhao Lu$^{1}$,
Yicheng Li$^{1}$,
Xian-Min Meng$^{2}$,
Xingchen Zhou$^{2}$
and Zuhui Fan$^{4}$
\\
$^{1}$Shanghai Key Lab for Astrophysics, Shanghai Normal University, Shanghai 200234, China\\
$^{2}$Key Laboratory of Space Astronomy and Technology, National Astronomical Observatories, Chinese Academy of Sciences,\\
20A Datun Road, Beĳing 100101, China\\
$^{3}$Science Center for China Space Station Telescope, National Astronomical Observatories, Chinese Academy of Sciences,\\ 
20A Datun Road, Beĳing 100101, China\\
$^{4}$South-Western Institute for Astronomy Research, Yunnan University, Kunming 650500, China\\
}

\date{Accepted XXX. Received YYY; in original form ZZZ}

\pubyear{2015}

\begin{document}
\label{firstpage}
\pagerange{\pageref{firstpage}--\pageref{lastpage}}
\maketitle

\begin{abstract}
  Accurate photometric redshift (photo-$z$) estimation requires support from multi-band observational data. However, in the actual process of astronomical observations and data processing, some sources may have missing observational data in certain bands for various reasons. This could greatly affect the accuracy and reliability of photo-$z$ estimation for these sources, and even render some estimation methods unusable. The same situation may exist for the upcoming Chinese Space Station Telescope (CSST). In this study, we employ a deep learning method called Generative Adversarial Imputation Networks (GAIN) to impute the missing photometric data in CSST, aiming to reduce the impact of data missing on photo-$z$ estimation and improve estimation accuracy. Our results demonstrate that using the GAIN technique can effectively fill in the missing photometric data in CSST. Particularly, when the data missing rate is below 30\%, the imputation of photometric data exhibits high accuracy, with higher accuracy in the $g$, $r$, $i$, $z$, and $y$ bands compared to the $NUV$ and $u$ bands. After filling in the missing values, the quality of photo-$z$ estimation obtained by the widely used Easy and Accurate Zphot from Yale (EAZY) software is notably enhanced. Evaluation metrics for assessing the quality of photo-$z$ estimation, including the catastrophic outlier fraction ($f_{out}$), the normalized median absolute deviation ($\rm {\sigma_{NMAD}}$), and the bias of photometric redshift ($bias$), all show some degree of improvement. Our research will help maximize the utilization of observational data and provide a new method for handling sample missing values for applications that require complete photometry data to produce results.

\end{abstract}




\section{Introduction}

The measurement of galaxy redshift plays a crucial role in understanding the large-scale structure of the universe, as well as the formation and evolution of galaxies \citep{tasca2009zcosmos,mo2010galaxy,conselice2014evolution}. By systematically detecting and analyzing the redshift of galaxies, their distances can be accurately determined, laying the foundation for studying various physical properties of galaxies, including mass, luminosity, and the scale of various extreme phenomena. Moreover, measuring galaxy redshift allows for a deeper investigation into e.g. the structure, origin, and evolution of the observable universe as a whole \citep{abdalla2011comparison,zhou2021spectroscopic}.

In general, the precise redshift of galaxies can usually be obtained by analyzing their spectral features, which is known as spectroscopic redshift (spec-$z$) \citep{carrasco2013tpz,cole2005power,percival2007shape}. However, spectroscopy is limited by its requirement for high wavelength resolution, often necessitating long exposure and integration times to achieve high signal-to-noise ratio spectra \citep{salvato2019many}. In recent years, a more efficient but less accurate statistical method for redshift estimation based on broad-band photometry, known as photometric redshifts (photo-$z$), has attracted increasing attention \citep[e.g.][]{koo1985optical,loh1986photometric,wolf2001multi,arnouts2002measuring,collister2004ANNz,babbedge2004IMPZ,feldmann2006zurich,rowan2008photometric,hildebrandt2008blind,coupon2009photometric,soo2018morpho}.The adoption of photo-$z$ enables cosmological measurements on galaxy samples that are orders of magnitude larger than comparable spectroscopic samples in the observable universe \citep[e.g.][]{colless2001spectra,dickinson2003great,Skrutskie2006two,garilli2008vimos,garilli2014vipers,Wright2010wise,Grogin_2011_candles,Parkinson2012wigglez,chambers2016panstarrs,aihara2017hsc,zou2019photometric}. These photo-$z$ samples typically have simpler and more uniform selection functions, extending to fainter flux limits and larger angular scales, thereby probing significantly larger cosmic volumes  \citep{hildebrandt2010phat}. Many ongoing and upcoming ground-based and space telescopes, including the Sloan Digital Sky Survey (SDSS) \citep{fukugita1996sloan,york2000sloan}, Dark Energy Survey (DES) \citep{collaboration2016more,abbott2021dark}, the Legacy Survey of Space and Time (LSST) \citep{ivezic2019lsst,abell2009lsst}, the Euclid Space Telescope \citep{laureijs2011euclid}, the Wide-Field Infrared Survey Telescope or Nancy Grace Roman Space Telescope(WFIRST) \citep{spergel2015widefield,green2012wide,akeson2019wide}, Multi-Object Optical and Near-infrared Spectrograph (MOONS) \citep{cirasuolo2020moons,maiolino2020moonrise}, 4-metre Multi-Object Spectroscopic Telescope (4MOST) \citep{de20194most}, and others, rely on high-precision photo-$z$ estimation to achieve their scientific goals. 

The accuracy of estimating photo-$z$ is affected by various factors, including the quality of photometric data, the coverage range of photometric bands, and the selection of redshift estimation methods, among others \citep{newman2022photometric}. Among these factors, the coverage range of photometric bands is considered particularly critical. Keeping the redshift estimation method and the quality of photometric data constant, utilizing photometric data with broader band coverage can improve the precision of photo-$z$ estimation. This is attributed to the fact that a wider band coverage enables the capture of more spectral information and provides additional data points for analysis, leading to more accurate redshift determinations \citep{salvato2019many}.

For example, the photo-$z$s obtained in the COSMOS2015 catalogue \citep{laigle2016cosmos2015} have been shown to be more precise and accurate compared to surveys like the Dark Energy Survey (DES) and the Sloan Digital Sky Survey (SDSS) \citep{gunn20062}. This improvement can be attributed to the fact that the COSMOS photo-$z$s were computed using more than 30 bands that cover a wide range of the electromagnetic spectrum, in contrast to the limited four or five optical bands utilized in the DES and SDSS surveys. 

Although leveraging photometric data from more bands can enhance the accuracy of photo-$z$ estimation, it is inevitable that many sources will have missing observations in one or more bands during practical observations in large-scale astronomical surveys due to various limitations. These factors comprise galaxies or regions that have not been observed in a specific band, galaxies or regions that are masked, photometric measurements that do not meet the detection threshold of the catalog, or photometry measurements characterized by a notably low signal-to-noise ratio \citep{euclid2023selection}.The lack of such data not only greatly affects the accuracy and reliability of photo-$z$ estimation, but also makes some estimation methods unusable due to incomplete photometry data. For instance, many machine learning algorithms developed for photo-$z$ estimation usually need full data from multiple bands as input \citep{lu2024estimating,fotopoulou2018cpz,zhou2021spectroscopic}. Hence, it is crucial to tackle the problem of missing data in observed photometry to maximize the utility of survey data and to achieve precise photo-$z$ estimation for these sources.

Various photo-$z$ estimation methods handle missing band data differently. For template fitting methods, missing bands are typically disregarded during redshift estimation \citep{bolzonella2011hyperz,arnouts2002measuring,ilbert2006accurate,brammer2008eazy}. For example, in the context of EAZY \citep{brammer2008eazy},If data is missing for a specific filter of an object during redshift calculation, the flux value for that band is typically set to a value more negative than the expected negative flux values, such as -99. This practice is undertaken to guarantee that this value is below any truly measured negative flux, as EAZY inherently manages non-detections with negative fluxes. Furthermore, EAZY also excludes objects in the catalog that have fewer filters than a specified number to uphold data reliability and precision.

When it comes to machine learning algorithms, a prevalent method to address missing data involves imputing values using techniques that are unrelated to the reasons for the data being absent. In the context of tree-based learning algorithms, various strategies can be employed to handle missing values. One common approach is to allocate a predefined constant value, such as -99.9, to signify the absence of data. Alternatively, missing values can be substituted with the mean, median, or minimum value of the corresponding feature, calculated across the galaxy sample using only the available data points. The selection of the imputation technique is guided by the specific demands of the algorithm and the nature of the dataset \citep{fotopoulou2018cpz,mucesh2021machine,schirmer2022euclid}. In neural network-based learning algorithms, missing values are generally not allowed, and complete data from multiple bands are required as input. Objects with photometric SEDs that are missing one or more bands are typically discarded from the analysis \citep{zhou2021spectroscopic, zhou2022photometricBNN}. This highlights the importance of data completeness for neural network applications.

While these methods for handling missing data work well in some photo-$z$ estimation algorithms, such as in tree-based CPz (Classification-aided photometric redshift estimation) algorithm, where substituting with the mean produces good outcomes, or in certain galaxy classification applications like selecting quiescent galaxies based on methods by \citet{euclid2023selection}, it is worth noting that these imputation methods for missing data do not provide close-to-real predictive values. Therefore, in other methods that require complete multi-band photometry data for accurate photo-$z$ estimation, the applicability of galaxies with missing data is limited by these imputation methods.

With the increasing popularity of machine learning, various machine learning methods for data imputation have emerged in the literature \citep{van2011mice,pereira2020reviewing,yoon2018gain,shang2017vigan,lee2019collagan}. These methods are also gradually being utilized to address the issue of missing data in the field of astronomy \citep{ren2020using,pichara2013automatic,keerin2022estimation,luken2021missing}. In this study, we employed a deep learning method based on Generative Adversarial Imputation Networks (GAIN) \citep{yoon2018gain} to impute missing data across one or multiple observed wavelength bands in a large sample of galaxies. This method has been tested in a small radio continuum catalogue and proven effective \citep{luken2021missing}. To the best of our knowledge, this is the first time that this method has been applied to photo-$z$ estimation, with the exception of \citet{luken2021missing}. Our aim is to ensure that the imputed data align closely with the actual observational values, thus reducing the influence of missing data on photo-$z$ estimation and enhancing the precision of the estimates.

\begin{figure}
	\includegraphics[width=0.9\columnwidth]{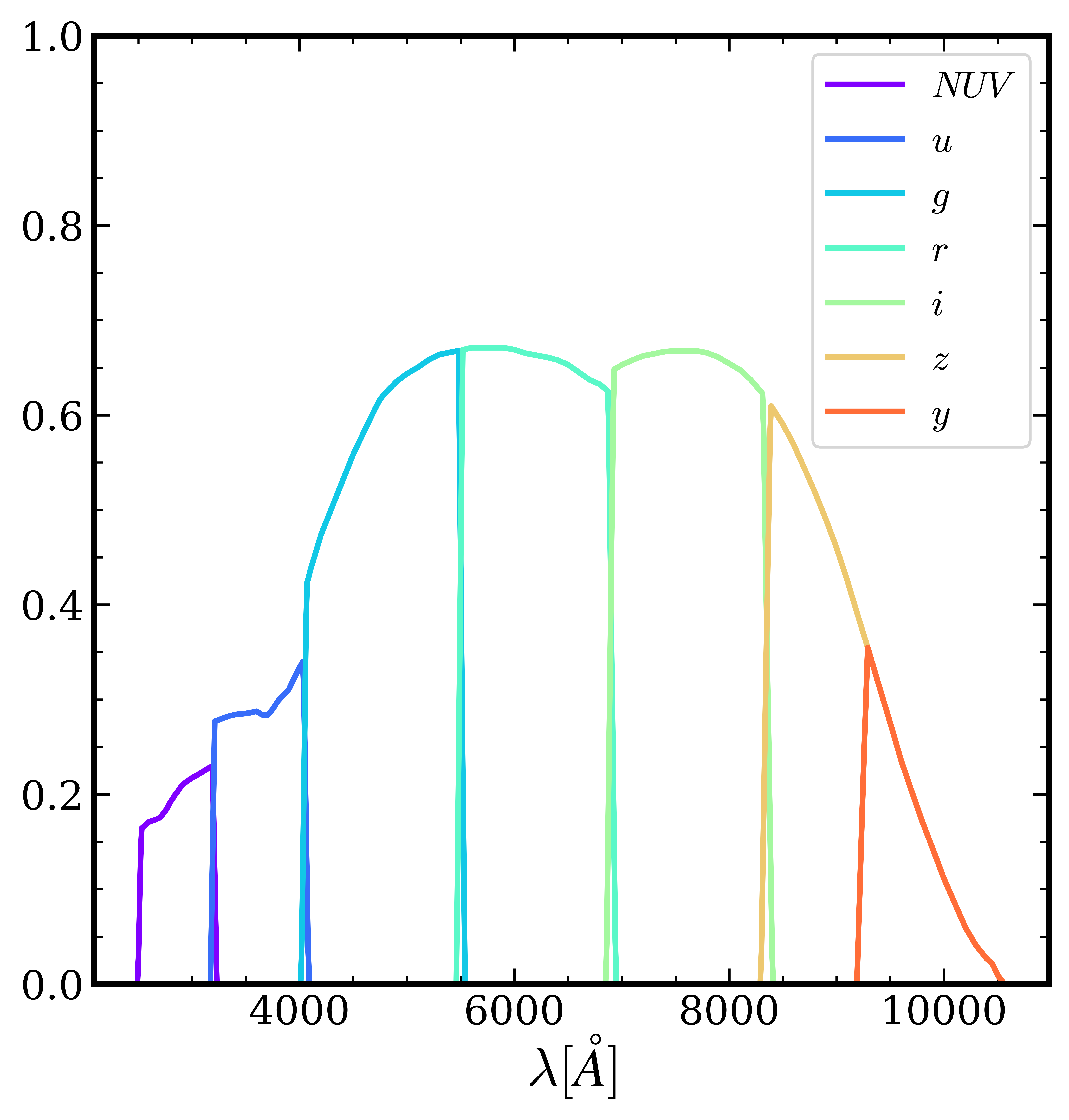}
	\caption{The solid curves illustrate the transmissions of the seven photometric bands used in CSST. These curves take into account the impact of detector quantum efficiency. For more information on the specific transmission parameters, please refer to \citet{cao2018testing}}
	\label{fig:filters}
\end{figure}

We implemented the GAIN method on simulated optical survey data obtained from the Chinese Space Station Telescope (CSST). Anticipated to be launched within next two years, the CSST will share the orbit with the China Manned Space Station \citep{zhan2011consideration,cao2018testing,gong2019cosmology}. The CSST survey is designed to cover approximately 17,500 deg$^2$ over a period of approximately 10 years, encompassing the optical and near-infrared (NIR) bands from approximately 250 nm to 1000 nm. The $5\sigma$ limit for point source magnitudes can reach around 26 AB mag for the $g$, $r$, and $i$ bands, and approximately 24.5 to 25.5 for the other bands. Figure \ref{fig:filters} illustrates the transmissions that are under test, including detector quantum efficiency, of the seven photometric filters employed by CSST. The primary scientific goals of CSST involve investigating the evolution of large-scale structure, the properties of dark matter and dark energy, as well as galaxy formation and evolution, among others \citep{gong2019cosmology,cao2022calibrating,zhan2021wide}. Thus, accurate photo-$z$ measurements are essential for achieving these objectives.

We rigorously assessed the accuracy of the imputed photometric values and then conducted photo-$z$ estimation using template fitting. By comparing the changes in photo-$z$ estimation accuracy before and after imputation, we evaluated the effectiveness of the missing data imputation method we employed. It is important to note that after imputation using our proposed method, galaxies with missing values can seamlessly be utilized in various photo-$z$ estimation methods or other applications, just like galaxies with complete data.

The organization of this paper is as follows. In Section \ref{section:mock}, we explain the process used to generate the mock photometry catalogues utilized in this study. Section \ref{section:GAIN} provides a detailed description of our adopted missing value imputation method and presents the results obtained from applying the method to the CSST mock missing samples. In Section \ref{section:photoz}, we compare the accuracy of photo-$z$ obtained using the EAZY method before and after imputation. Finally, we summarize the results and draw conclusions in Section \ref{section:conclusions}.

\section{mock data}\label{section:mock}

In this section, we will provide a brief overview of how the mock data is generated. More detailed information on the mock process can be found in \citet{zhou2021spectroscopic}. This catalog has been employed in several studies aligned with CSST objectives, with a focus on exploring machine learning and spectral energy distribution fitting techniques. The mock data is designed to have similar characteristics as the observations from the CSST survey, including redshift, magnitude distribution, and galaxy types. To ensure a high level of realism in simulating galaxy images for the CSST photometric survey, we utilize mock image generation techniques based on observations taken within the COSMOS field using the Advanced Camera for Surveys of the Hubble Space Telescope (HST-ACS), while incorporating CSST instrumental effects. The mock flux data of galaxies are measured from these images using aperture photometry. The COSMOS HST-ACS survey encompasses an area of approximately 2 deg$^2$ in the F814W band, which has a spatial resolution similar to that of the CSST, with an 80\% energy concentration radius of $R_{80}\sim0.^{\prime\prime}15$ \citep{cao2018testing,gong2019cosmology,koekemoer2007cosmos,massey2010pixel,bohlin2016perfecting}. Moreover, the COSMOS HST-ACS F814W survey exhibits significantly lower background noise compared to the CSST survey, expected to be approximately 1/3 of that experienced in the CSST survey. This attribute provides a solid foundation for the simulation of CSST galaxy images. For more comprehensive information on the mock data generation process, we recommend referring to \citet{zhou2021spectroscopic}. Here, we summarize and rewrite the important points.

First, we select an area of 0.85 × 0.85 deg$^2$ from the HST ACS survey, where $\sim$ 192,000 galaxies can be identified. Then we rescale the pixel size from $0.^{\prime\prime}03$ of the HST survey to $0.^{\prime\prime}075$ of the CSST survey. The identified galaxies are extracted as square stamp images with galaxies at the centers of images. The image sizes are 15 times the semi-major axis of galaxies, which can be obtained in the COSMOS weak lensing source catalog \citep{leauthaud2007weak}, so our galaxy images have different sizes. Other sources in the image are masked and replaced by background noise, and only the galaxy image in the center is preserved.

Next, we proceed to rescale the galaxy images from the HST-ACS F814W survey to the CSST flux level. This is done by utilizing galaxy spectral energy distributions (SEDs) to obtain the CSST 7-band images. The galaxy SEDs are generated by fitting the fluxes and other photometric information provided in the COSMOS2015 catalog using the LePhare code \citep{arnouts1999measuring,ilbert2006accurate,laigle2016cosmos2015}. During this fitting process, the photo-$z$s from the catalog are fixed. The SED templates used for fitting are also sourced from this catalog, and they are extended from  $\sim$ 900Å to $\sim$ 90Å using the BC03 method \citep{bruzual2003stellar}. This extension allows for the inclusion of fluxes from high-redshift galaxies in all CSST photometric bands. Further details can be found in the work of \citet{cao2018testing}.

We select around 100,000 high-quality galaxies with reliable photo-$z$ measurements for the SED fitting process. In addition to dust extinction, we also consider emission lines such as Ly$\alpha$, H$\alpha$, H$\beta$, [O\uppercase\expandafter{\romannumeral2}], and [O\uppercase\expandafter{\romannumeral3}]. After fitting the galaxy SEDs, we can calculate the theoretical flux data by convolving them with the CSST filter transmission curves, as depicted in Figure \ref{fig:filters}. Simultaneously, we calculate the fluxes of the F814W images using an aperture size of 2 times the Kron radius \citep{kron1980photometry}. The CSST 7-band images are then produced by rescaling the fluxes accordingly. To match the CSST observations, the background noise is also adjusted to the same level. Further details regarding the noise adjustment can be found in \citet{zhou2022photometricBNN}. As a result, we obtain the mock CSST galaxy images for the seven CSST photometric bands.

To measure the flux in our galaxy mock data, we employ aperture photometry. Initially, we determine the Kron radius along the major and minor axes, allowing us to define an elliptical aperture size of 1 times $R_{Kron}$. Within this aperture, the flux and its corresponding error can be calculated for each band. 

The final CSST mock catalog comprises measurements in seven bands of the CSST, including flux, flux error, and photo-$z$, for nearly 100,000 galaxies. These CSST mock galaxies were selected from the COSMOS catalog, which utilizes photo-$z$ estimates computed from over 30 bands covering a wide range of the electromagnetic spectrum. \citet{laigle2016cosmos2015} conducted a verification process by comparing photo-$z$ estimates in the COSMOS2015 catalog with various spectroscopic survey samples. Additional information regarding the accuracy of the photo-$z$ estimates and the characteristics of spectroscopic redshift (spec-$z$) samples can be found in Tables 4 and 5, as well as Figures 11 and 12 presented by \citet{laigle2016cosmos2015}. Based on the demonstrated precision and accuracy of the photo-$z$ estimates in the COSMOS catalog, we consider them reliable and have adopted them as the true redshift values, referred to as $z_{\rm true}$ hereafter. 

From this CSST simulated catalog, we further selected a high-quality photometric sub-sample, where the signal-to-noise ratio (SNR) is greater than 10 in either the $g$ or $i$ band, with valid observations in other bands.

The sub-sample obtained in this way is referred to as the High-quality CSST sample (HCS), which includes 40,763 sources.

\begin{figure}
	\includegraphics[width=\columnwidth]{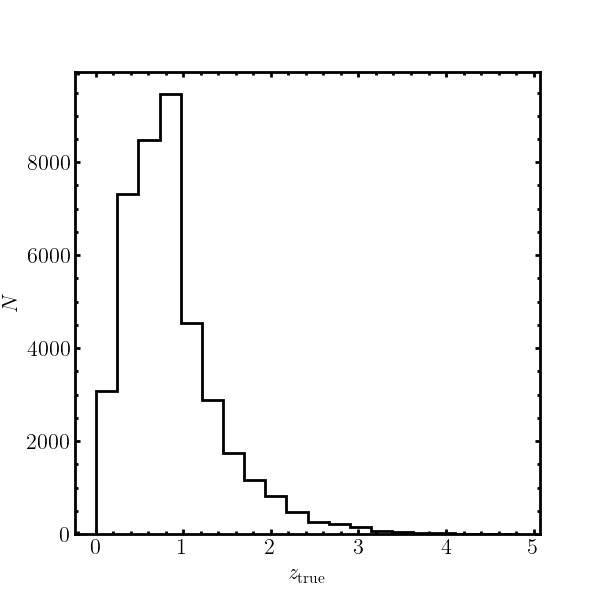}
	\caption{The galaxy redshift distribution of the HCS dataset. The distribution peaks around $z = 0.8 \sim 1.0$, and can reach maximum at $z \sim 5$.}
	\label{fig:redshift}
\end{figure}

\begin{figure}
	\includegraphics[width=\columnwidth]{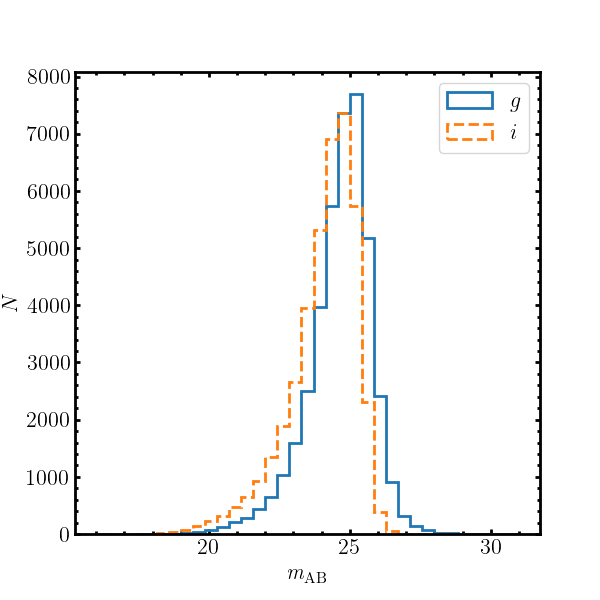}
	\caption{the distribution of AB magnitudes in the $g$ and $i$ bands for the HCS dataset.}
	\label{fig:mag_g_i}
\end{figure}

Figure \ref{fig:redshift} illustrates the distribution of redshifts for the HCS dataset. The plot shows that the majority of sources have redshifts concentrated around the range of $z=0.8-1.0$. The redshift distribution extends from 0 to 5, implying the inclusion of sources spanning a wide range of cosmic distances. 
Figure \ref{fig:mag_g_i} provides the distribution of AB magnitudes in the $g$ and $i$ bands for the HCS dataset. This plot gives an overview of the brightness distribution in these specific bands for the selected high-quality sub-sample.

 Furthermore, we simulated different missing rates to create multiple sub-samples for imputation testing. The missing rate is defined as $n/7$, where $n$ is the number of missing bands. For each source in HCS, we randomly deleted values from one band, resulting in a sub-sample with a missing rate of approximately $14.3\%$, known as the One-band Missing Sample (OMS). Similarly, we also randomly removed 2, 3, or 4 different bands for each source in HCS, creating three additional sub-samples referred to as the Two-band Missing Sample (TMS), Three-band Missing Sample (TrMS), and Four-band Missing Sample (FMS). Each source in these sub-samples has 2, 3, or 4 missing values, with missing rates approximately $28.6\%$, $42.9\%$, and $57.1\%$ respectively.

\section{Imputation of Missing Data} \label{section:GAIN}

\subsection{Missing Data}

Missing values are frequently encountered in astronomical datasets, with various underlying causes. These include: 1) incomplete observations, such as galaxies or regions not observed in specific wavelength bands, as well as galaxies or regions that are masked.  2) instrument constraints and adverse observing conditions, leading to photometric measurements that fall below catalog detection thresholds or have notably low signal-to-noise ratios. 3) recording errors and data corruption. 4) missing values resulting from combining multiple survey data, for example, due to variations in depth between different surveys or the absence of certain objects in a particular survey.

Based on the missingness assumptions, the issue of missing data can be classified into three categories: missing completely at random (MCAR), missing at random (MAR), and missing not at random (MNAR) \citep{little2019statistical,graham2009missing,van2018flexible}. In astronomical research, missing data can also be classified into these three categories. Specifically, MCAR indicates that missing data is random and unrelated to any observed or unobserved factors; MAR suggests that missing data is related to observed variables but not to unobserved variables; MNAR indicates that missing data is related to unobserved factors. In the majority of astronomical research samples, most missing data is typically categorized as either MCAR or MAR, with MNAR being less common.

For data that are MCAR or MAR, many imputation algorithms can be utilized to estimate missing values based on existing observed/measured data. An effective imputation algorithm should not only accurately predict missing values but also maintain the consistency and accuracy of feature-label relationships after imputation.

State-of-the-art imputation methods can be classified into two categories: discriminative methods and generative methods. Discriminative methods focus on modeling the relationship between observed and missing data directly, often using statistical techniques or machine learning algorithms to predict missing values based on other variables. On the other hand, generative methods aim to model the joint distribution of observed and missing data, creating a complete representation of the data distribution. These methods use approaches such as Bayesian networks or latent variable models to impute missing values by considering the underlying structure of the data. Discriminative methods encompass MICE \citep{white2011multiple}, MissForest \citep{stekhoven2012missforest}, and matrix completion techniques, while generative methods involve expectation-maximization algorithms and deep learning models like denoising autoencoders (DAE) and generative adversarial networks (GAN). \citet{yoon2018gain} introduced a novel approach that leverages the well-known GAN framework for filling missing data. They termed this method Generative Adversarial Imputation Nets (GAIN). Through evaluating GAIN's performance on diverse datasets, they discovered that GAIN significantly outperforms other state-of-the-art imputation methods in terms of imputation accuracy in substituting MCAR data. Further research has shown that the GAIN method also performs well on MAR data \citep{dong2021generative}.

Next, we provide a detailed description of the GAIN method used in this study and present the results obtained by applying the GAIN method to mock missing samples from CSST.

\subsection{GAIN Algorithm}

The GAIN algorithm is based on the well-known GAN framework and is designed to handle the unique characteristics of the imputation problem. Various experiments conducted on real-world datasets demonstrate that GAIN outperforms current state-of-the-art imputation techniques. For a more comprehensive understanding, we highly recommend referring to the research paper by \citet{yoon2018gain}. Here, we will provide a general overview of the GAIN algorithm.

The architecture of GAIN is shown in Figure 4. It consists of two main components: a generator ($G$) and a discriminator ($D$). The $G$ takes in incomplete data as input and aims to accurately impute the missing values. It does so by generating plausible values for the missing data based on the observed data and any available contextual information. On the other hand, the $D$'s role is to distinguish between the observed (complete) data and the imputed data generated by the $G$. It is trained to minimize the classification loss, which involves classifying whether each component was observed or imputed. Both the $G$ and $D$ are trained simultaneously in an adversarial process. The $G$'s training objective is to maximize the $D$'s misclassification rate, while the $D$'s objective is to correctly classify the observed and imputed components. This adversarial training process ensures that the $G$ improves its imputation performance over time, gradually generating more accurate imputations. The architecture of GAIN builds upon and adapts the standard GAN architecture to handle the specific challenges and characteristics of the missing data imputation problem. 

\begin{figure*}
	\includegraphics[width=0.7\textwidth]{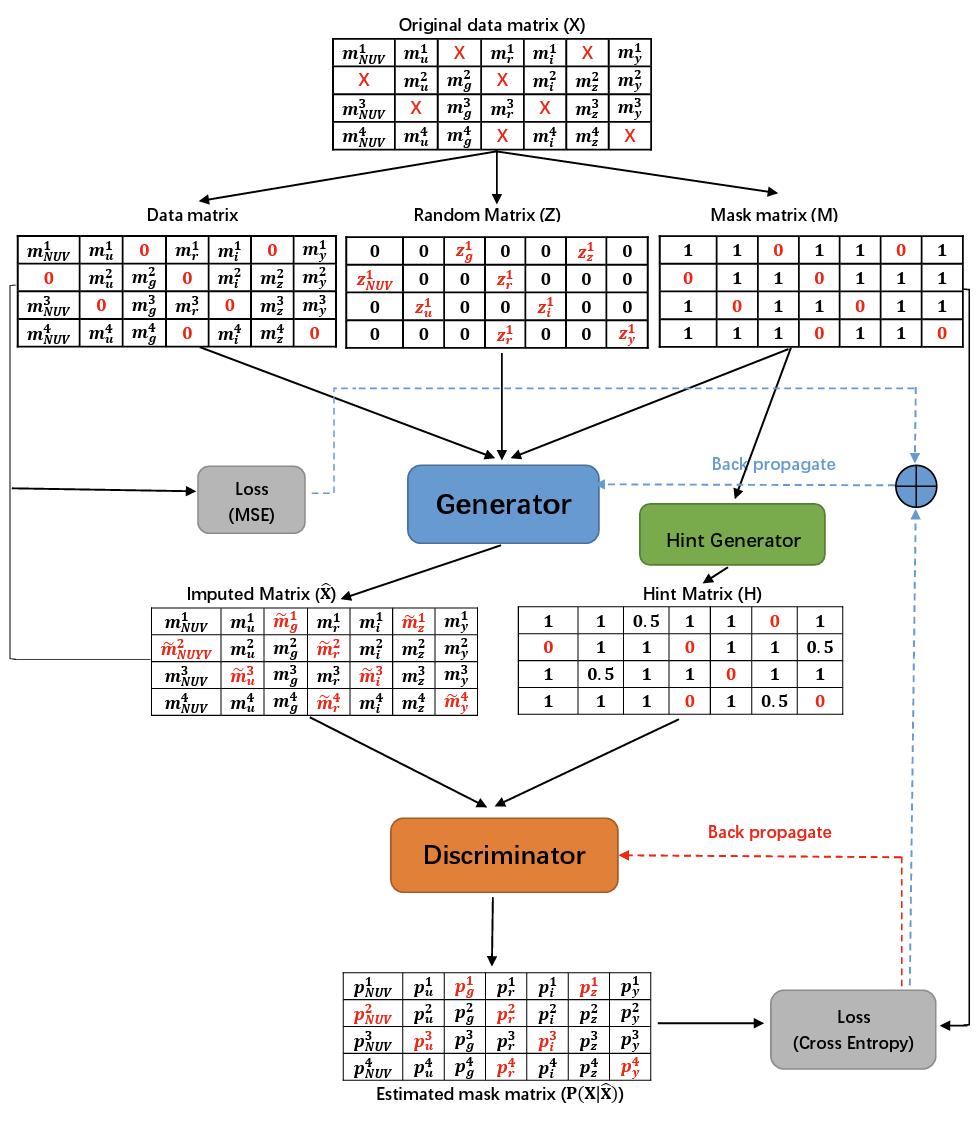}
	\caption{The architecture of GAIN. All the subscripts denote the CSST filter, and the superscripts represent the index of the galaxy sample. "$m$" represents the true observed magnitude of the galaxies, "$\widetilde{m}$" represents the predicted magnitude after imputation, "$z$" denotes the random noise, "$p$" represents the probability distribution. Additionally, the symbol "X" in red represents the missing magnitudes in that particular filter.}
	\label{fig:GAIN_archi}
\end{figure*}

The detailed imputation procedures with GAIN can be summarized as follows. Firstly, let's consider a dataset X that requires imputation. Here, M is a mask matrix of the same shape as X, used to identify the missing values. Additionally, Z is a matrix of the same shape as X, representing random noise that is independent of all other variables. To address potential issues with different variable scales affecting variable weights, we standardize all variables in X to a range of 0-1. Next, a fully connected multilayer neural network generator $G$ is constructed. $G$ takes X, M, and Z as inputs. During the initial training phase, the output of $G$ can be represented as:
\begin{equation}
    \bar{X}=G((\mathrm{X} \odot \mathrm{M}+\mathrm{Z} \odot(1-\mathrm{M})), \mathrm{M}),
	\label{eq:output_G}
\end{equation}
The imputed matrix, denoted as $\hat{X}$, can be defined as:
\begin{equation}
    \hat{X}=\mathrm{X} \odot \mathrm{M}+\bar{X} \odot(1-\mathrm{M}).
	\label{eq:imputed_matrix}
\end{equation}
A fully connected multilayer neural network discriminator $D$ is then constructed. $D$ takes the imputed dataset generated by the generator and a hint matrix H as inputs. The hint matrix H is created using a hint mechanism to enhance discrimination:
\begin{equation}
    \mathrm{H}=\mathrm{B} \odot \mathrm{M}+0.5(1-\mathrm{B}).
	\label{eq:hint_matrix}
\end{equation}
Here, B is a random variable: 
\begin{equation}
    \mathrm{B}=\left(\mathrm{B}_1, \ldots, \mathrm{B}_d\right) \in\{0,1\}^d.
	\label{eq:B_var}
\end{equation}
The discriminator D outputs the probability of a value being real or fake:
\begin{equation}
    D_{\text {output }}=D(\widehat{X}, \mathrm{H}).
	\label{eq:D_output}
\end{equation}
Next, we construct adversarial networks and start training. We randomly extract a mini-batch of k samples from the dataset for the training process. Training begins by optimizing the loss function for $D$, followed by alternating training between $D$ and $G$, optimizing their respective loss functions:
\begin{equation}
    D_{\text {loss }}=\max _D[\mathrm{M} \log D(\hat{X}, \mathrm{H})+(1-\mathrm{M}) \log (1-D(\hat{X}, \mathrm{H}))],
	\label{eq:D_loss}
\end{equation}
\begin{equation}
    G_{\text {loss }}=\min _G\left(L_G+\alpha L_M\right),
	\label{eq:G_loss}
\end{equation}
and
\begin{equation}
    L_G=-(1-\mathrm{M}) \log D(\hat{X}, \mathrm{H}).
	\label{eq:gain_loss}
\end{equation}
Here, $L_M$ represents the sum of mean squared errors (MSE) between the observed values and the predicted values at the corresponding locations, and $\alpha$ is a hyper-parameter. Finally, we apply the inverse standardization process to restore the imputed values back to their original scale.

\label{sec:maths} 

\subsection{Imputation Results on CSST Mock Data} \label{subsection:experiments}

In this section, we assess the imputation performance of GAIN using four datasets: OMS, TMS, TrMS and FMS. In all experiments, the GAIN model is configured with a 6-layer generator and discriminator. The number of hidden nodes in each layer is 14, 14, 7, 7, 7 and 7, respectively. The activation function used for each layer, except for the output layer, is $tanh$. The output layer employs the $sigmoid$ activation function. The number of batches is set to 256 for both the generator and discriminator. The hyperparameter $\alpha$ is set to 100, and the hint rate is set to 0.85.

The accuracy of imputation was evaluated by measuring the imputation error, which is defined as the difference between the imputed values and the actual values. In this study, the normalized root mean square error (NRMSE) was used as the assessment metric. NRMSE was defined as follows:
\begin{equation}
    \mathrm{NRMSE}=\frac{\sqrt{\frac{1}{N} \sum_{i=1}^N\left(\widehat{x}_i-x_i\right)^2}}{\frac{\sum_{i=1}^N x_i}{N}},
	\label{eq:nrmse}
\end{equation}
where $\hat{x}_i$ represents the imputed value and $x_i$ represents the original value, and $N$ is the total number of missing values.

The GAIN method was applied for imputation 100 times for each of the simulated incomplete datasets, OMS, TMS, TrMS, and FMS, with each imputation involving a complete retraining. The NRMSE values obtained from these 100 imputations were analyzed, and the average NRMSE and standard deviation were calculated.

\begin{table*}
	\centering
	\caption{Imputation errors (NRMSE) on the OMS, TMS, TrMS and FMS datasets.}
	\label{tab:NRMSE}
    \resizebox{\textwidth}{!}{
		\begin{tabular}{|c|c|c|c|c|c|c|c|c} 
		\hline
    & $NUV$          & $u$     & $g$     & $r$     & $i$     & $z$     & $y$     & all bands   \\ \hline
OMS & 0.040 $\pm$ 0.008 & 0.022 $\pm$ 0.007 & 0.012 $\pm$ 0.003 & 0.010 $\pm$ 0.001 & 0.007 $\pm$ 0.002 & 0.007 $\pm$ 0.002 & 0.010 $\pm$ 0.006 & 0.021 $\pm$ 0.005 \\ \hline
TMS & 0.050 $\pm$ 0.042 & 0.025 $\pm$ 0.013 & 0.016 $\pm$ 0.013 & 0.012 $\pm$ 0.013 & 0.009 $\pm$ 0.004 & 0.009 $\pm$ 0.011 & 0.011 $\pm$ 0.004 & 0.025 $\pm$ 0.017 \\ \hline
TrMS & 0.077 $\pm$ 0.059 & 0.052 $\pm$ 0.056 & 0.032 $\pm$ 0.045 & 0.028 $\pm$ 0.043 & 0.021 $\pm$ 0.035 & 0.027 $\pm$ 0.041 & 0.029 $\pm$ 0.036 & 0.051 $\pm$ 0.037 \\ \hline
FMS & 0.185 $\pm$ 0.079 & 0.187 $\pm$ 0.095 & 0.131 $\pm$ 0.070 & 0.099 $\pm$ 0.068 & 0.091 $\pm$ 0.061 & 0.090 $\pm$ 0.052 & 0.105 $\pm$ 0.063 & 0.144 $\pm$ 0.054 \\ \hline
	\end{tabular}
    }
\end{table*}

Table \ref{tab:NRMSE} shows the imputation errors (NRMSE) and their standard deviations obtained from using the GAIN method to impute missing values in the OMS, TMS, TrMS, and FMS datasets. According to the data in the table, it can be observed that for the OMS and TMS datasets, which have relatively low missing rates (missing rate $<30\%$), the GAIN method performs well and yields small imputation errors.  This suggests that when one or two bands of data are missing in CSST photometry data, the GAIN method proposed in this study can accurately restore the missing values. However, as the data's missing rate gradually increases, the imputation error of the GAIN method also increases. For the TrMS and FMS datasets, with missing rates of 42.9\% and 57.1\% respectively, which are higher than the OMS and TMS datasets, the imputation errors for missing data in each band are significantly increase.

\begin{figure}
	\includegraphics[width=\columnwidth]{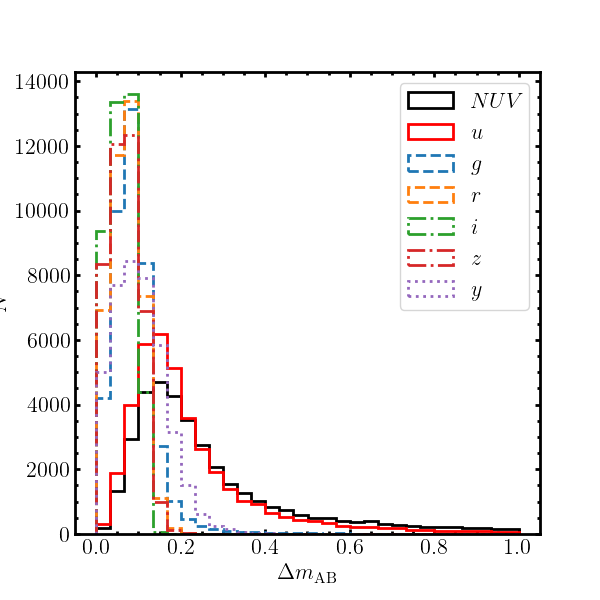}
	\caption{The apparent magnitude error distribution for each band in our simulated CSST data (HCS).}
	\label{fig:mag_err}
\end{figure}

Furthermore, from our research, we found that among the seven bands of CSST, when imputing missing values using the GAIN method, the $NUV$ band has the largest imputation errors, followed by the $u$ band, while the $g$, $r$, $i$, $z$, and $y$ bands show very good imputation performance, especially at lower missing rates. Apart from the model itself, we believe that the accuracy of CSST photometry data may also have an impact on the results. In the CSST photometry data, the $NUV$ band has the least accuracy, followed by the $u$ band, while the accuracy of the $g$, $r$, $i$, $z$, and $y$ bands is the highest. This suggests that larger photometry data errors may introduce certain uncertainties to the model training, leading to larger imputation uncertainties when filling missing values.  Figure \ref{fig:mag_err} displays the distribution of photometry data errors for each band in our CSST simulated data (HCS). From the figure, it can be observed that both the mean and variance of the error distributions for the $NUV$ and $u$ bands are significantly higher than those for the other bands.

\begin{figure*}
	\includegraphics[width=\textwidth]{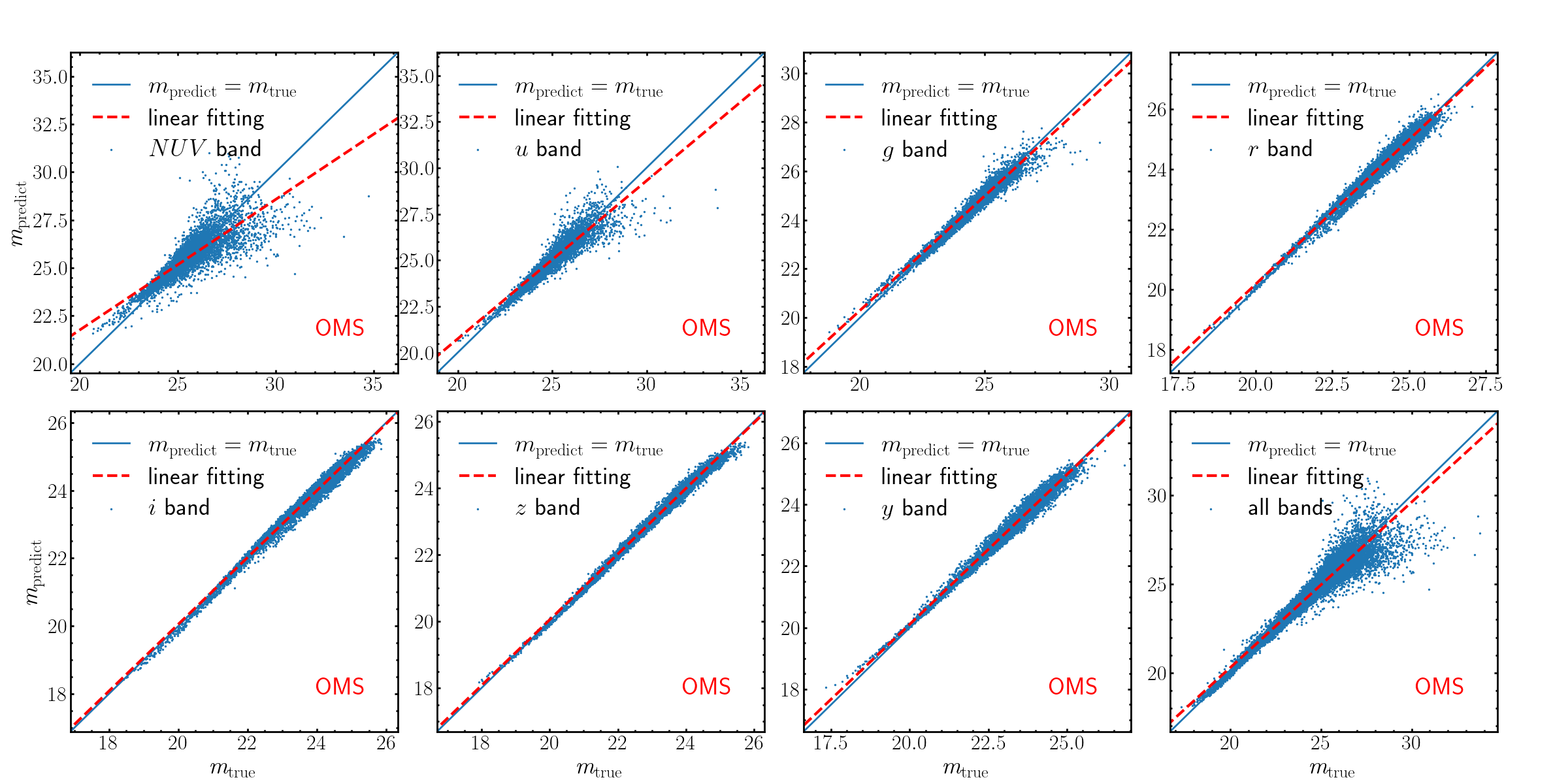}
	\caption{Imputed values vs. True values for the OMS dataset: The x-axis represents the true photometric values of the missing bands in the OMS dataset, while the y-axis represents the predicted values obtained using the GAIN method. The solid line represents $m_\mathrm{predict} = m_\mathrm{true}$}, and the dashed line represents the results of linear regression fitting for all data points in the subplot.
	\label{fig:diff_OMS}
\end{figure*}

\begin{figure*}
	\includegraphics[width=\textwidth]{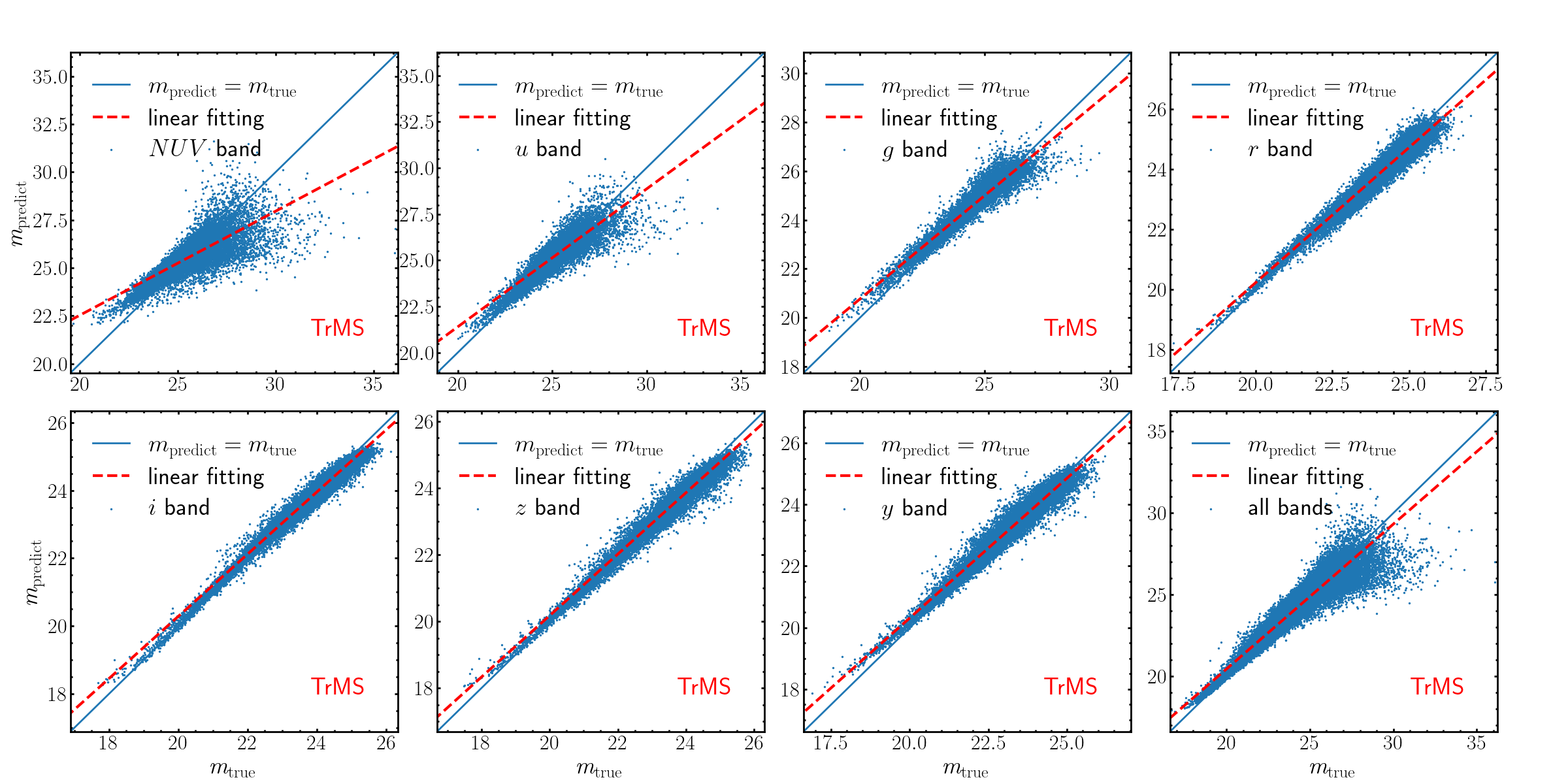}
	\caption{Imputed values vs. True values for the TrMS dataset: The x-axis represents the true photometric values of the missing bands in the FMS dataset, while the y-axis represents the predicted values obtained using the GAIN method. Both solid and dashed lines have the same meaning as in figure \ref{fig:diff_OMS}.
}
	\label{fig:diff_TrMS}
\end{figure*}

After imputing missing data 100 times, we take the average of the imputed values as the predicted value for the missing data, and consider the standard deviation as the error of the predicted value. Figures \ref{fig:diff_OMS} shows the comparison between the true values and the predicted values for the OMS dataset. The solid line represents $m_\mathrm{predict} = m_\mathrm{true}$, and the dashed line represents the result of linear regression fitting to the data points in the subplot. It can be clearly seen that the GAIN method is effective in imputing missing photometry data, especially at lower missing rates. Similarly, in the TMS dataset, the GAIN method also performs well. However, in cases of higher missing rates such as the TrMS and FMS datasets, the imputation accuracy of the GAIN method significantly decreases. Figure \ref{fig:diff_TrMS} shows the comparison between the predicted values and true values in the TrMS dataset, where the missing rate reaches $57.1\%$. By comparing Figure \ref{fig:diff_OMS} and Figure \ref{fig:diff_TrMS}, the impact of missing rates on imputation performance can be clearly observed. The lower the missing rate, the better the imputation effect of the GAIN method on the photometry data.

\begin{figure*}
	\includegraphics[width=\textwidth]{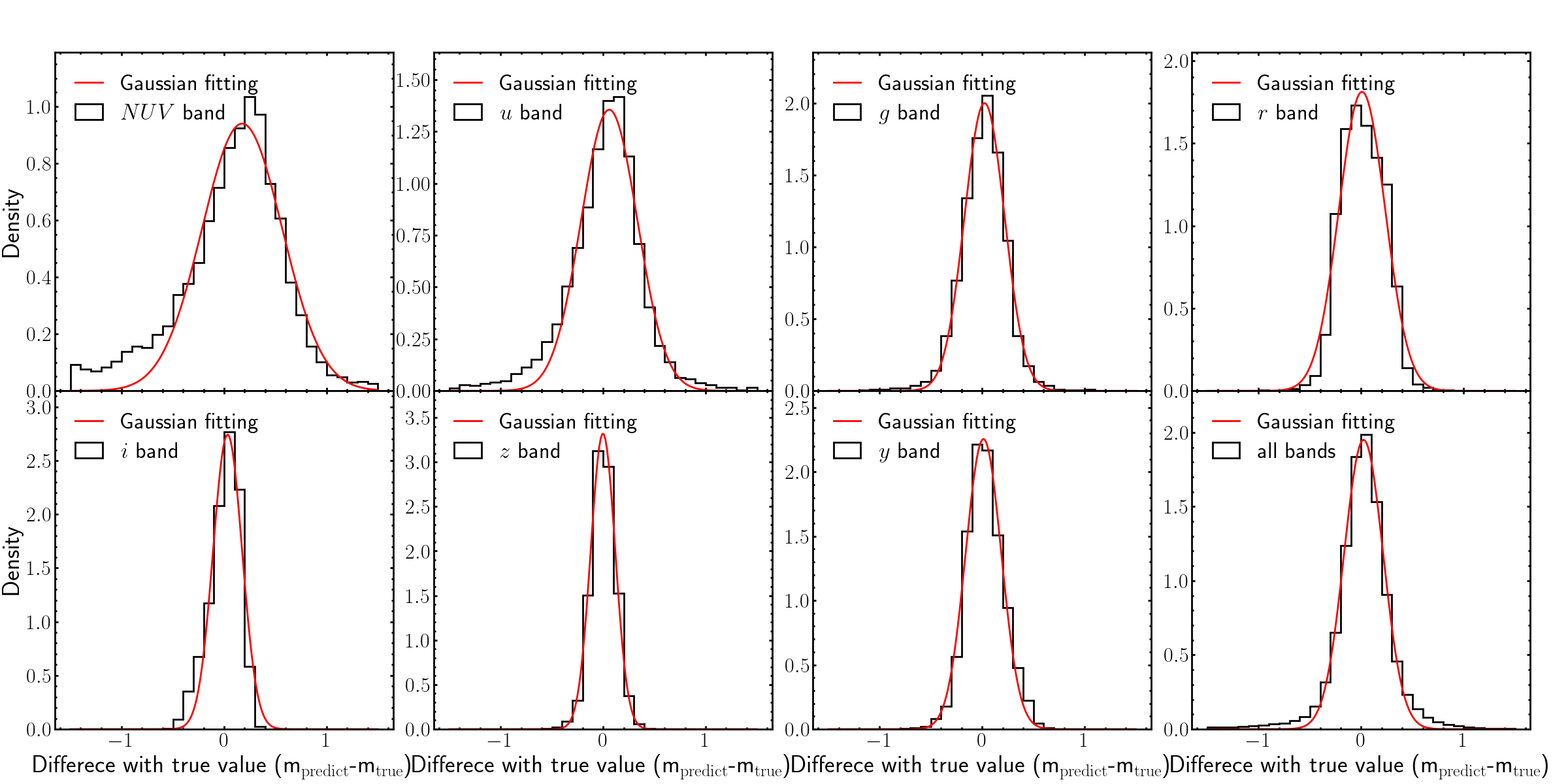}
	\caption{Density plots illustrating the distribution of discrepancies between the imputed values and true values for each band in the OMS dataset. The black line represents the distribution of differences between imputed and true values, while the red line indicates the Gaussian fit of this distribution.
}
	\label{fig:diff_dist_OMS}
\end{figure*}

In order to compare the differences between imputed values and true values across different bands, we utilized density plots to visualize the filling status of each band within the OMS dataset. In Figure \ref{fig:diff_dist_OMS}, the distribution of discrepancies between the predicted values by GAIN and the true values for missing data in each band is displayed by the black line, along with the Gaussian fit represented by the red line. It can be observed from the graph that the differences between the filling values and the true values in each band are close to zero and concentrated, indicating high filling accuracy. However, the $NUV$ and $u$ bands tend to exhibit a broader distribution of differences and a higher density of larger differences. Moreover, our research findings suggest that the differences between filling values and true values become more pronounced in datasets with higher missing rates, such as TMS, TrMS, and FMS.

\begin{figure*}
	\includegraphics[width=\textwidth]{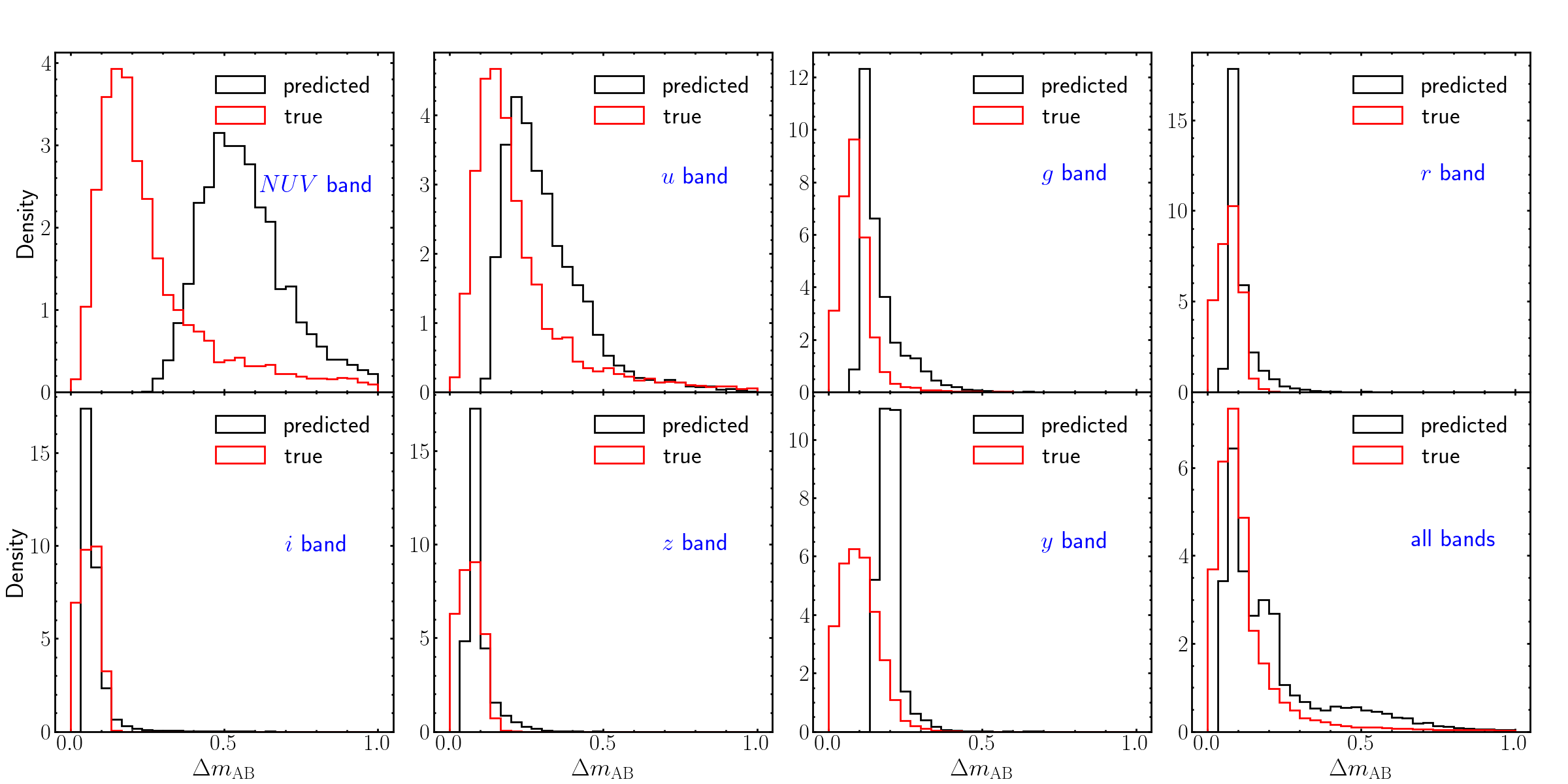}
	\caption{The error distribution of predicted values (black line) and true values (red line) for each band in the OMS dataset. The last subplot (bottom right) represents the overall error distribution of all bands.
}
	\label{fig:err_OMS}
\end{figure*}

Finally, we compared the error distribution of predicted values with that of true values. Through this comparison, we can comprehensively evaluate the accuracy and reliability of the model, understand the degree of difference between predicted results and true values, and provide guidance for further model improvement. In Figure \ref{fig:err_OMS}, we illustrate the error distribution of predicted values (black line) and true values (red line) in all datasets. It can be observed from the graph that, in all bands, the errors of our model's predicted values are larger than those of the true values, especially in the $NUV$ and $u$ bands , which is an expected and consistent outcome. Nevertheless, as illustrated in the lower right subplot, the overall error distribution of predicted values closely corresponds with that of true values.

\section{Applications in Photo-$z$ Estimation} \label{section:photoz}

In the previous section, missing values were filled in the simulated datasets (OMS, TMS, TrMS, and FMS) of CSST. In this section, the EAZY template fitting method will be utilized to estimate the photo-$z$s of these datasets before and after imputation. The necessity of missing data imputation and the effectiveness of the GAIN method will be validated by comparing the alterations in the quality metrics of photo-$z$ estimation for the samples.

\subsection{Photo-$z$ Quality Metrics}

To assess the quality of our photo-$z$ estimates on the sample, we introduce three metrics. The first metric is the normalized median absolute deviation (NMAD), which measures accuracy and is defined as \citep{brammer2008eazy}
\begin{equation}
	{\rm \sigma_{NMAD}}=1.48\times {\rm median}\left(\left|\frac{\Delta z - {\rm median}(\Delta z)}{1+z_{\rm true}}\right|\right),
\end{equation}
 where 
\begin{equation}
	\Delta z =z_{\rm phot}-z_{\rm true},
\end{equation}
$z_{\rm true}$ is the reference redshift used as the ‘ground truth’ and $z_{\rm phot}$ is the predicted photo-$z$. NMAD is preferred over standard deviation as it is less sensitive to outliers and incorporates a scaling factor of 1.48, allowing NMAD to be interpreted as the standard deviation for normally distributed data. 

The second metric is the proportion of catastrophic outliers ($f_{out}$). Sources whose photo-$z$ estimates satisfy the following condition \citep{fotopoulou2018cpz,euclid2023selection},
\begin{equation}
	\frac{|z_{\rm true}-z_{\rm phot}|}{1+z_{\rm true}}>0.15,
\end{equation}
are considered catastrophic outliers, meaning their photo-$z$s are incorrect. 

The final metric is the bias of the photometric redshifts ($bias$), denoted as
\begin{equation}
	bias = {\rm median}\left(\frac{z_{\rm true}-z_{\rm phot}}{1+z_{\rm true}}\right),
\end{equation}
which examines whether and to what extent we systematically overestimate or underestimate the redshifts of the galaxies.

\subsection{Photo-$z$ Results}

There are many methods available for estimating redshift from photometric data, mainly divided into template fitting and machine learning methods (for details, see the review by \citet{salvato2019many}). Among template fitting methods, EAZY is a commonly used software. For example, \citet{yang2014photometric} used EAZY to estimate the redshift of the Hawaii-Hubble Deep Field-North (H-HDF-N) survey catalog, while \citet{chen2018xmm} utilized EAZY to estimate the photo-$z$s of X-ray point sources in the XMM-Large Scale Structure (XMM-LSS) survey region. Both studies demonstrated the excellent performance of EAZY. In the upcoming tests, we will use EAZY to estimate the photo-$z$s of our samples. According to the study by \citet{desprez2020euclid}, when run with the same configuration, different template fitting methods can provide almost identical results. The observed differences are not due to differences in the performance of template fitting methods, but rather due to differences in their configurations. Therefore, we expect that our test results will also hold true for other template fitting methods.

In our EAZY analysis, we made use of the standard CWW+KIN template set, which is based on the CWW empirical template set \citep{2003Colors}  with the extension prescribed by \citet{Kinney1996Template}. This template set includes six templates and is commonly employed in photo-$z$ estimation. Additionally,  we incorporated the $r$ band apparent magnitude prior $p(z|m_r)$, which represents the redshift distribution of galaxies with the apparent magnitude $m_r$. The parameter $Z\_STEP\_TYPE$ was set to 0, indicating a uniformly spaced redshift grid. Other parameters were kept at their default settings. The input bands included the seven CSST bands.

\begin{table*}
    \centering
    \caption{ Quality metrics of photo-$z$ estimation for HCS, OMS, TMS, TrMS, and FMS datasets before and after imputation using EAZY with $r$ band prior.}
    \label{tab:photoz_rp}
    \resizebox{\textwidth}{!}{
    \begin{tabular}{|c|c|cc|cc|cc|cc|}
    \hline
                 & HCS & \multicolumn{2}{c|}{OMS}     & \multicolumn{2}{c|}{TMS}     & \multicolumn{2}{c|}{TrMS}    & \multicolumn{2}{c|}{FMS}     \\ \cline{2-10} 
                 & No Missing & \multicolumn{1}{c|}{\begin{tabular}[c]{@{}c@{}}Before\\ Imputation\end{tabular}} & \begin{tabular}[c]{@{}c@{}}After\\Imputation\end{tabular} & \multicolumn{1}{c|}{\begin{tabular}[c]{@{}c@{}}Before\\Imputation\end{tabular}} & \begin{tabular}[c]{@{}c@{}}After\\Imputation\end{tabular} & \multicolumn{1}{c|}{\begin{tabular}[c]{@{}c@{}}Before\\Imputation\end{tabular}} & \begin{tabular}[c]{@{}c@{}}After\\Imputation\end{tabular} & \multicolumn{1}{c|}{\begin{tabular}[c]{@{}c@{}}Before\\Imputation\end{tabular}} & \begin{tabular}[c]{@{}c@{}}After\\Imputation\end{tabular} \\ \hline
${\rm \sigma_{NMAD}}$                 & 0.039 & \multicolumn{1}{c|}{0.048}  & 0.047  & \multicolumn{1}{c|}{0.065}  & 0.061  & \multicolumn{1}{c|}{0.099}  & 0.083  & \multicolumn{1}{c|}{0.187}  & 0.133  \\ \hline
$f_{out}$                & 3.56\% & \multicolumn{1}{c|}{6.43\%}  & 6.50\%  & \multicolumn{1}{c|}{12.54\%}  & 11.02\%  & \multicolumn{1}{c|}{23.86\%}  & 17.96\%  & \multicolumn{1}{c|}{43.63\%}  & 31.20\%  \\ \hline
$bias$                & 0.0047 & \multicolumn{1}{c|}{0.0027}  & 0.0017  & \multicolumn{1}{c|}{0.0068}  & 0.0023  & \multicolumn{1}{c|}{0.0268}  & 0.0061  & \multicolumn{1}{c|}{0.0098}  & 0.0141  \\ \hline
    \end{tabular}
    }
\end{table*}

\begin{table*}
    \centering
    \caption{Quality metrics of photo-$z$ estimation for HCS, OMS, TMS, TrMS, and FMS datasets before and after imputation using EAZY without prior.}
    \label{tab:photoz_np}
    \resizebox{\textwidth}{!}{
    \begin{tabular}{|c|c|cc|cc|cc|cc|}
    \hline
& HCS & \multicolumn{2}{c|}{OMS}     & \multicolumn{2}{c|}{TMS}     & \multicolumn{2}{c|}{TrMS}    & \multicolumn{2}{c|}{FMS}     \\ \cline{2-10} 
                  & No Missing & \multicolumn{1}{c|}{\begin{tabular}[c]{@{}c@{}}Before\\Imputation\end{tabular}} & \begin{tabular}[c]{@{}c@{}}After\\Imputation\end{tabular} & \multicolumn{1}{c|}{\begin{tabular}[c]{@{}c@{}}Before\\Imputation\end{tabular}} & \begin{tabular}[c]{@{}c@{}}After\\Imputation\end{tabular} & \multicolumn{1}{c|}{\begin{tabular}[c]{@{}c@{}}Before\\Imputation\end{tabular}} & \begin{tabular}[c]{@{}c@{}}After\\Imputation\end{tabular} & \multicolumn{1}{c|}{\begin{tabular}[c]{@{}c@{}}Before\\Imputation\end{tabular}} & \begin{tabular}[c]{@{}c@{}}After\\Imputation\end{tabular} \\ \hline
${\rm \sigma_{NMAD}}$                 & 0.047 & \multicolumn{1}{c|}{0.060}  & 0.057  & \multicolumn{1}{c|}{0.089}  & 0.080  & \multicolumn{1}{c|}{0.175}  & 0.158  & \multicolumn{1}{c|}{0.435}  & 0.425  \\ \hline
$f_{out}$                 & 7.94\% & \multicolumn{1}{c|}{12.52\%}  & 11.07\%  & \multicolumn{1}{c|}{22.48\%}  & 18.64\%  & \multicolumn{1}{c|}{41.52\%}  & 38.49\%  & \multicolumn{1}{c|}{67.15\%}  & 65.77\%  \\ \hline
$bias$& 0.0916 & \multicolumn{1}{c|}{0.1361}  & 0.1041  & \multicolumn{1}{c|}{0.2259}  & 0.1586  & \multicolumn{1}{c|}{0.3883}  & 0.3599  & \multicolumn{1}{c|}{0.5913}  & 0.5887  \\ \hline
    \end{tabular}
    }
\end{table*}

Upon utilizing the EAZY software and incorporating the $r$-band apparent magnitude prior, the photo-$z$ estimation results for all samples in this research are presented in Table \ref{tab:photoz_rp}. The term "Before imputation" denotes the results of samples prior to data imputation, whereas "After imputation" indicates the results after filling in the missing data. It can be seen from Table 2 that the quality of photo-$z$ estimation has significantly improved overall after missing value imputation. As the level of missing data in the sample increases (from OMS->TMS->TrMS->FMS), the enhancement in the accuracy of photo-$z$ estimation after imputing missing values becomes more noticeable compared to pre-imputation. In the case of OMS, there are only minimal changes in the three metrics assessing photo-$z$ estimation quality. However, from TMS to TrMS and then to FMS, there is a substantial improvement in these metrics. The catastrophic outlier fraction ($f_{out}$) has shown relative improvements of 12.1\% (TMS), 24.7\% (TrMS), and 28.5\% (FMS) after imputation. Furthermore, both the normalized median absolute deviation ($\sigma_\mathrm{NMAD}$) and the bias of photometric redshift ($bias$) have also significantly improved. Importantly, in scenarios with higher rates of missing data (such as TrMS and FMS), utilizing imputed data for photo-$z$ estimation has demonstrated increased accuracy and practical value compared to the pre-imputation results.

On the other hand, the results of using EAZY for photo-$z$ calculation without incorporating prior information are shown in Table \ref{tab:photoz_np}. From the table we can see that after imputing all missing values in the samples, the three metrics evaluating the quality of photo-$z$ estimation have also improved. However, unlike Table \ref{tab:photoz_rp}, the improvement in the quality of photo-$z$ estimation is not monotonically increasing with the increasing proportion of missing data in the samples. For example, from OMS to TMS and then to TrMS and finally to FMS, the improvement rate of the catastrophic outlier fraction $f_{out}$ shows a monotonically increasing trend in Table \ref{tab:photoz_rp}, while in Table \ref{tab:photoz_np}, the improvement rates of $f_{out}$ are not monotonically increasing. More specifically, without prior information in all datasets, TMS shows the most significant improvement in the quality of photo-$z$ estimation after filling in the missing values, with the catastrophic outlier rate $f_{out}$ improving by approximately 17.1\%, the normalized median absolute deviation ($\sigma_\mathrm{NMAD}$) improving by about 11.3\%, and the photometric redshift bias ($bias$) improving by 29.8\%. The OMS dataset follows, with $f_{out}$ improving by 11.6\%, $\sigma_\mathrm{NMAD}$ improving by 23.5\%, and $bias$ improving by 6\%. Conversely, the TrMS and FMS datasets with higher rates of missing data show a smaller improvement in metrics after imputation, and their actual utility is limited due to the lower accuracy of photo-$z$ estimation in these samples.

By considering the results in Section \ref{subsection:experiments}, we can understand the reasons for these differences. In the GAIN method, the accuracy of imputation for the $r$ band is relatively high even with a high data missing rate. Therefore, by incorporating prior information of the $r$ band in the EAZY algorithm, the results of photo-$z$ estimation can be significantly improved, even in the TrMS and FMS datasets, as shown in Table \ref{tab:photoz_rp}. 

Upon comparing Table \ref{tab:photoz_rp} and \ref{tab:photoz_np}, we can also see that for samples containing missing data, such as OMS, TMS, TrMS, and FMS, either adding prior information to the r-band or imputing missing values can enhance the accuracy of photo-$z$ estimation. However, the improvement in accuracy of photo-$z$ is more significant when both adding prior information to the $r$-band and imputing missing values simultaneously. The enhancement from adding $r$-band prior information without imputing missing values can be observed in the "Before Imputation" columns of Table \ref{tab:photoz_rp} and \ref{tab:photoz_np}, while the improvement from imputing missing values without adding prior information can be seen in the "Before Imputation" and "After Imputation" columns of Table \ref{tab:photoz_np}. The increase in accuracy of photo-$z$ after imputing missing values and adding $r$-band prior information can be observed in the "Before Imputation " and "After Imputation" columns of Table \ref{tab:photoz_rp}. 

In  addition, we further tested the addition of prior information from other bands, such as the $i$-band, and found that this can also improve the photo-$z$ results for each sample. However, compared to adding prior information to the $r$-band, the improvement in photo-$z$ after adding it to other bands is less significant. This is because within the redshift range of our samples, the importance of all bands other than NUV in photo-$z$ estimation is lower than that of the $r$-band, as illustrated in Figure 13 of \citet{lu2024estimating}. Additionally, the photometric accuracy of the $NUV$ band is the lowest, making it unsuitable for constructing prior information. Therefore, in the photo-$z$ tests presented in this paper, we considered adding prior information specifically for the $r$-band.

\section{Conclusions} \label{section:conclusions}

In this study, we employed a deep learning technique known as Generative Adversarial Imputation Networks (GAIN) \citep{yoon2018gain} to impute missing photometric data in CSST. Our study will help improve the utilization of CSST observational data in the future and provide a new alternative method for handling missing data in large observation samples. Although our study focuses on CSST, this method can also be applied to ongoing or upcoming surveys such as LSST, Euclid, DES, and others. By following the outlined imputation method in this paper, datasets with missing values can be efficiently integrated into various software applications.

The CSST survey includes photometry in seven bands, namely $NUV$, $u$, $g$, $r$, $i$, $z$, and $y$, spanning optical and near-infrared wavelengths. Photometric observations for CSST were simulated using data from HST-ACS and the COSMOS catalog, taking into account the instrument effects specific to CSST. Mock galaxy images in seven bands were generated, and flux along with observational error data were calculated using photometric apertures. Initial samples for this study were selected based on sources with signal-to-noise ratios exceeding 10 in the $g$ or $i$ bands and valid observations present in all bands. Subsequently, photometric data for all seven CSST bands of each galaxy in this sample was randomly eliminated to create sub-samples with varying data missing rates.

Upon examining these sub-samples with missing data, we discovered that the GAIN method is effective in filling in the absent photometric data. Specifically, when the data missing rate is below 30\% in the samples, the accuracy of imputing the photometric data is notably high. Further comprehensive research indicates that among the seven observational bands of CSST, the $g$, $r$, $i$, $z$, and $y$ bands display the highest imputation accuracy, followed by the $u$ band, with the $NUV$ band exhibiting the lowest imputation accuracy. We believe that, in addition to factors related to the model itself, the poor quality of photometric data in the $NUV$ band of CSST may also impact the training of the GAIN model, leading to a significant decrease in imputation accuracy for missing values in the $NUV$ band.

After filling all missing samples using the GAIN method, we further utilized the template fitting software EAZY to obtain the photo-$z$s of the samples before and after imputation, and calculated three indicators to evaluate the quality of photo-$z$ estimation, including catastrophic outlier fraction ($f_{out}$), normalized median absolute deviation ($\sigma_{NMAD}$), and photometric redshift bias ($bias$). By comparing the changes in these three indicators before and after imputation, we further verified the effectiveness of the GAIN method proposed in this study. Detailed research findings show that regardless of whether prior information is added in the EAZY software, imputing missing values significantly improves the quality of photo-$z$ estimation. In particular, with the inclusion of prior information on the $r$-band, the improvement in photo-$z$ quality increases with the proportion of missing data in the samples. Furthermore, in cases of high data missing rates (>30\%), the photo-$z$s obtained after imputing missing data exhibit significant enhancement compared to before imputation, making rough photo-$z$ estimation possible for such galaxies. With higher missing rates, such as those observed in TrMS and FMS sub-samples, the imputation errors tend to be more pronounced. Despite the increased errors, it is important to note that the photo-$z$ estimation shows a more substantial improvement in these instances. This suggests that even with the inherent challenges posed by high missing rates, the accuracy of photo-$z$ estimation can benefit significantly from effective imputation techniques.

\section*{Acknowledgements}

ZJL acknowledges the support from the Shanghai Science and 
Technology Foundation Fund under grant No. 20070502400, and 
the science research grants from the China Manned Space Project. LPF acknowledges the support from the Innovation Program 2019-01-07-00-02-E00032 of SMEC. WD acknowledges the support from NSFC grant No. 11890691. YG acknowledges the support from National Key R\&D Program of China grant Nos. 2022YFF0503404, 2020SKA0110402, the CAS Project for Young Scientists in Basic Research (No. YSBR-092), and China Manned Space Project with Grant No. CMS- CSST-2021-B01. S.Z. acknowledges support from the National Natural Science Foundation of China (Grant No. NSFC-12173026), the Program for Professor of Special Appointment (Eastern Scholar) at Shanghai Institutions of Higher Learning and the Shuguang Program of Shanghai Education Development Foundation and Shanghai Municipal Education Commission. ZF acknowledges the support from NSFC grant No. U1931210. This work is also supported by the National Natural Science Foundation of China under Grants Nos. 12141302 and 11933002, and the science research grants from China Manned Space Project with Grand No. CMS-CSST-2021-A01. 

\section*{Data Availability}
 
The data that support the findings of this study are available from the corresponding author, upon reasonable request.



\bibliographystyle{mnras}
\bibliography{ref} 



\bsp	
\label{lastpage}
\end{document}